\documentclass[twocolumn,
amsmath,amssymb,aps,pra]{revtex4-1}

\usepackage{graphicx}
\usepackage{physics}
\usepackage{comment}
\usepackage{color}
\usepackage{dcolumn}
\usepackage{bm} 

\usepackage[colorlinks,urlcolor=blue,citecolor=blue,linkcolor=blue]{hyperref}

\newcommand{\p}{{\bf p}}

\newcommand{\q}{{\bf q}}
\renewcommand{\k}{{\bf k}}

\newcommand{\s}{{\bf s}}

\newcommand{\eb}{\varepsilon_{\rm X}}
\newcommand{\ep}{\epsilon}

\newcommand{\beq}{\begin{equation}}
\newcommand{\eeq}{\end{equation}}

\newcommand{\Frac}[2]{\displaystyle\frac{#1}{#2}}

\newcommand{\ch}{\hat{c}}
\newcommand{\chd}{\hat{c}^{\dagger}}
\newcommand{\e}{\hat{e}}
\newcommand{\ed}{\hat{e}^{\dagger}}
\newcommand{\h}{\hat{h}}
\newcommand{\hd}{\hat{h}^{\dagger}}

\newcommand{\w}{\omega}
\newcommand{\sg}{\sigma}

\newcommand{\replyadd}[1]{{\color{blue}#1}}

\begin{document}

\title{Trion resonance in polariton-electron scattering}

\author{Sangeet S. Kumar}
\affiliation{School of Physics and Astronomy, Monash University, Victoria 3800, Australia}
\affiliation{ARC Centre of Excellence in Future Low-Energy Electronics Technologies, Monash University, Victoria 3800, Australia}

\author{Brendan C. Mulkerin}
\affiliation{School of Physics and Astronomy, Monash University, Victoria 3800, Australia}
\affiliation{ARC Centre of Excellence in Future Low-Energy Electronics Technologies, Monash University, Victoria 3800, Australia}

\author{Meera M. Parish}
\affiliation{School of Physics and Astronomy, Monash University, Victoria 3800, Australia}
\affiliation{ARC Centre of Excellence in Future Low-Energy Electronics Technologies, Monash University, Victoria 3800, Australia}

\author{Jesper Levinsen}
\affiliation{School of Physics and Astronomy, Monash University, Victoria 3800, Australia}
\affiliation{ARC Centre of Excellence in Future Low-Energy Electronics Technologies, Monash University, Victoria 3800, Australia}

\date{\today}

\begin{abstract}
Strong interactions between charges and light-matter coupled quasiparticles offer an intriguing prospect with applications from optoelectronics to light-induced superconductivity. Here, we investigate how the interactions between electrons and exciton-polaritons in a two-dimensional semiconductor microcavity can be resonantly enhanced due to a strong coupling to a trion, i.e., an electron-exciton bound state. We develop a microscopic theory that uses a strongly screened interaction between charges to enable the summation of all possible diagrams in the polariton-electron scattering process, and we find that the polariton-electron interaction strength can be strongly varied and enhanced in the vicinity of the resonance. We furthermore derive an analytic approximation of the interaction strength based on universal low-energy scattering theory. This is found to match extremely well with our full calculation, indicating that the trion resonance is near universal, depending more on the strength of the light-matter coupling relative to the trion binding energy rather than on the details of the electronic interactions.
Thus, we expect the trion resonance in polariton-electron scattering to appear in a broad range of microcavity systems with few semiconductor layers, such as doped monolayer MoSe$_2$ where such resonances have recently been observed experimentally  [Sidler \textit{et al.}, Nature Physics \textbf{13}, 255 (2017)].

\end{abstract}

\maketitle

\section{Introduction}

Exciton polaritons are hybrid light-matter quasiparticles composed of a photon mode and an exciton (a bound electron-hole pair) in a semiconductor.  
Such quasiparticles can be created by placing a two-dimensional (2D) semiconductor layer in a microcavity, thereby enhancing the coupling between matter and light~\cite{QuantumFluidsofLight,Byrnes2014}. Due to their hybrid nature, polaritons have a very low mass inherited from their photon component, as well as the capability to interact with each other and other particles due to their matter component, which allows them to achieve condensation and superfluidity at elevated temperatures~\cite{Kasprzak2006,Amo2009,Sanvitto2010,Lerario2017}. The ability of polaritons to interact compared to ordinary photons leads to many applications such as ultra-fast polariton spin switching~\cite{Amo2010} and the emergence of photon correlations~\cite{Munoz2019,Delteil2019} with the potential prospect of realizing polariton blockade in a semiconductor device~\cite{Verger2006}.

\begin{figure}
\includegraphics[width=0.9\linewidth]{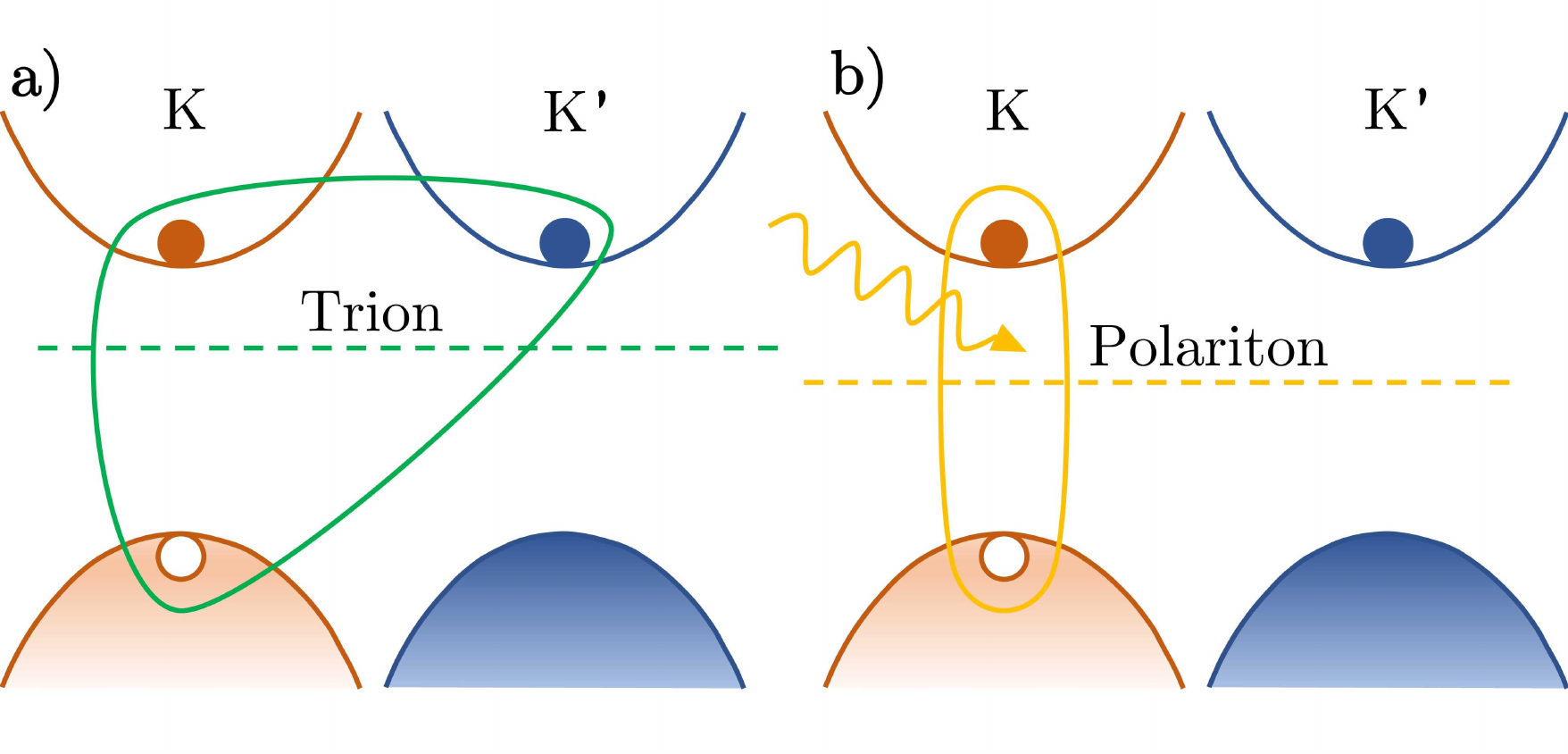}
\caption[system]{\label{TrionFig}
Schematic illustration of the trion resonance in intervalley polariton-electron scattering, using the  band structure of $\text{MoSe}_{2}$ as an example. Panel \textbf{a} shows the charges involved in the negatively charged trion (encircled by the green line), with the trion energy drawn in dashed green. In panel \textbf{b}, the exciton polariton formed by the photon and electron-hole pair in the K valley is encircled in yellow. The corresponding energy (dashed yellow) is tunable, allowing one to achieve a resonance condition with the trion. The electrons (holes) are represented by filled (empty) circles, and their spins are shown with red ($\sg=\uparrow$) and blue ($\sg=\downarrow$), corresponding to the K and K' valleys.
}
\end{figure}

In the past decade, a new class of 2D semiconductors has gained prominence, namely the monolayer transition metal dichalcogenides (TMDs). These have strong coupling to light, they can be externally tuned via electrostatic gating and doping techniques~\cite{WangRev2018} and, in addition to excitons, they feature trions (bound states of two electrons in distinct momentum-space valleys and a hole) that are potentially stable at room temperature~\cite{Jones2013,Mak2013}. These properties combine to make TMDs ideally suited for a broad range of applications in electronics and optoelectronics. Recently, it was demonstrated that TMDs can feature resonantly enhanced interactions between electrons and polaritons~\cite{SidlerNatPhys16}, as evidenced by a strongly doping-dependent optical response near the trion energy. The basic mechanism is illustrated in Fig.~\ref{TrionFig}: The coupling of light and matter allows the tuning of the energy of a polariton such that the total energy of a polariton and an electron matches that of the trion. The resulting coupling between these two configurations in a polariton-electron scattering process greatly enhances the interaction strength. Note that the same physics applies to polaritons and holes, but we focus on polariton-electron scattering to be concrete in the following. 

In this paper, we develop a diagrammatic description of polariton-electron scattering in a 2D semiconductor microcavity, taking into account both the strong coupling to light and the composite nature of the exciton and trion bound states. Our approach allows us to determine the sum of all diagrams contributing to the scattering process. We find a strong enhancement of the polariton-electron interaction strength for parameters corresponding to typical TMD monolayers, with the resonance occurring at slightly negative detuning. For systems with larger light-matter (Rabi)
coupling relative to the trion binding energy, such as GaAs semiconductor microcavities with multiple quantum wells, the resonance shifts to large positive detuning, making it inaccessible in typical experiments. This likely explains why the resonant enhancement has not previously been observed in III-V semiconductors. 

The electron-exciton scattering problem is challenging to solve even in the absence of a strong coupling to light~\cite{CombescotPRX2017,Fey2020,Efimkin2021}.
Therefore, to make the calculation tractable, we use a highly screened interaction between charges, as done previously for intravalley polariton-polariton and polariton-electron scattering~\cite{Li2021PRL,Li2021,Li2021PRBPP}. However, we argue that our results are dominated by the strong light-matter coupling, rather than the precise form of the electronic interactions. Indeed, inspired by Ref.~\cite{Bleu2020} we derive an analytic expression for the polariton-electron scattering which is based on only two assumptions: (i) that the polariton-electron scattering can be viewed as off-shell exciton-electron scattering, where the collision energy is determined by the light-matter coupling; and (ii) that the exciton-electron scattering is given by the universal low-energy scattering formula of short-range systems~\cite{AdhikariAJP86}. With these two assumptions, we arrive at the expression for the polariton-electron interaction strength:
\beq \label{AnalyticTintro}
  g_{\rm{eP}} \simeq X_{-}^{2} \Frac{2\pi}{m_{\rm{eX}}}
  \frac1
  {\ln[
  (E_\mathrm{T}+\eb)
  /(E_{-}+\eb) ]  },
\eeq  
where $\eb$ is the exciton binding energy, $m_{\rm{eX}}=(m_{\rm{e}}^{-1}+m_{\rm{X}}^{-1})^{-1}$ is the reduced mass of electrons and excitons (mass $m_{\rm{e}}$ and $m_{\rm{X}}$, respectively), $E_-$ and $X_-$ are the lower polariton energy and exciton amplitude at zero momentum, $E_\mathrm{T}$ is the trion energy, and energies are measured from that of the electron-hole continuum. Equation~\eqref{AnalyticTintro} illustrates the tunability of the interactions, since the polariton energy $E_-$ depends on the cavity mode frequency and thus the cavity length. Crucially, it provides a near-perfect agreement with our numerical results for TMD monolayers, thus highlighting the universal nature of our results. This is all the more remarkable, since the system of three charges features both a direct (bright) and an indirect (dark) exciton, with both of these configurations contributing to the trion wave function in our full theory. Yet, the analytic approximation Eq.~\eqref{AnalyticTintro} only considers the direct exciton. The key feature that underlies this universal behavior is the separation of energy scales between the exciton and the trion binding energies, with $\eb \gg |E_{\mathrm{T}}+\eb|$. 

The paper is organized as follows. In Sec.~\ref{TheoreticalFramework} we introduce the Hamiltonian and two-body theory of exciton-polaritons which forms the basis of the three-body problem. In Sec.~\ref{PolaritonElectronScattering}, we develop the intervalley polariton-electron scattering equation by iterating the irreducible three-body exchange process, and we show our results for parameters corresponding to a microcavity containing a TMD monolayer or a GaAs quantum well. In Sec.~\ref{ConcludingRemarks}, we summarize and provide a brief outlook.
Technical details are given in the appendices. 

\section{Theoretical framework} \label{TheoreticalFramework}
\subsection{Hamiltonian}

To describe polariton-electron scattering and its connection to the trion resonance, we consider a minimal microscopic model that encompasses all the necessary ingredients. That is, it features exciton binding, trion binding, and strong coupling to light. Specifically, we characterize the 2D semiconductor microcavity by an effective Hamiltonian that includes light, matter, and light-matter coupling:
\begin{align} \label{Hamiltonian}
    \hat H=\hat H_{\rm{ph}}+\hat H_{\rm{mat}}+\hat H_{\rm{ph-mat}}.
\end{align}

The matter part of the Hamiltonian $\hat{H}_{\rm{mat}}$ describes the single-particle energies and interactions of the electrons and holes, and is given by
\beq\label{eq:Ham}
\begin{split}
\hat{H}_{\rm{mat}} 
 = & \underset{\k\sg}{\sum} \bigg{(} 
\ep^{\rm{e}}_{\k} \ed_{\sg, \k} \e_{\sg, \k} +
\ep^{\rm{h}}_{\k} \hd_{\sg, \k} \h_{\sg, \k} \bigg{)} \\
& - V_0 \underset{
\stackrel{ \k \k^{'} \q }{ \sg \sg' }
}{\sum}  
 \ed_{\sg, \k+\q}\hd_{\sg', \k^{'}-\q}\h_{\sg', \k^{'}}\e_{\sg, \k} .
\end{split}
\eeq
Here, the creation (annihilation) operators of electrons and holes with momentum $\k$ and spin $\sg$ are  $\ed_{\sg, \k} \; (\e_{\sg, \k})$ and $\hd_{\sg, \k} \; (\h_{\sg, \k})$, with corresponding dispersions $\ep^{\rm{e/h}}_{\k} = \k^2 /2m_{\rm{e/h}}$ in terms of their effective masses $m_{\rm{e}}$ and $m_{\rm{h}}$. The single-particle energies are measured with respect to the band gap. Here, and in the following, we work in units where the area $\mathcal{A}=1$ and $\hbar=1$.

To describe the attractive interactions between electrons and holes, we use a highly screened contact interaction of strength $V_0 >0$, which is related to the exciton binding energy $\varepsilon_\mathrm{X}$ via
\begin{align}\label{eq:1overv}
    \frac1{V_0}=\sum_\k^\Lambda \frac1{\varepsilon_\mathrm{X}+\ep^{\rm{e}}_{\k}+\ep^{\rm{h}}_{\k}}.
\end{align}
Here, $\Lambda$ is an ultraviolet cutoff of order the inverse lattice spacing such that $\Lambda a_X \gg 1$, where we have defined the effective Bohr radius $a_\mathrm{X}$ via $\varepsilon_\mathrm{X}=1/2m_{\rm{r}} a_\mathrm{X}^2$, with $m_{\rm{r}} = (m_{\rm{e}}^{-1} + m_{\rm{h}}^{-1})^{-1}$ the reduced mass of the electron-hole pair. Since our Hamiltonian does not describe the high-energy physics of the system such as the details of the band structure, we will eventually take $\Lambda$ to infinity to obtain cutoff-independent results~\cite{LevinsenBook15}. In this limiting renormalization process, the interaction strength $V_0$ approaches 0 as $1/\ln\Lambda$ according to Eq.~\eqref{eq:1overv}. Note that we do not explicitly include the electron-electron and hole-hole repulsion, since highly screened repulsive interactions generally yield a much smaller scattering cross section than their attractive counterpart and can thus be neglected.
However, given that these are not required to obtain the exciton and trion bound states, this does not strongly impact our results. As we discuss below, the important consequence of electron-electron repulsion is that it impacts the ratio of the exciton to trion binding energies (since the trion features two electrons), which we instead adjust using a method borrowed from nuclear physics.

The photon part of the Hamiltonian consists of the cavity mode, which acquires an effective mass in the microcavity. It is given by
\beq
\hat{H}_{\rm{ph}} = \underset{\k\sg}{\sum} (\w + \ep^{\rm{c}}_{\k} )\chd_{\sg, \k} \ch_{\sg, \k} .
\eeq
Here, the operators $\chd_{\sg, \k}$ and $\ch_{\sg, \k}$, respectively, create and annihilate a microcavity photon with momentum $\k$ and polarization $\sg$, with corresponding kinetic energy  $\ep^{\rm{c}}_{\k} = \k^2/2m_{\rm{c}}$. Throughout, we take the effective mass of the cavity photon to be $m_{\rm{c}}=2\times10^{-4}m_{\rm r}$~\cite{QuantumFluidsofLight}. For simplicity, we write the zero-momentum bare cavity photon energy $\omega$ separately, noting that this is also measured from the electronic band gap.

Lastly, the light-matter component of the Hamiltonian $\hat{H}_{\rm{ph-mat}}$ describes the transformation of a microcavity photon to an intravalley electron-hole pair, and vice versa. Importantly, due to the optical selection rules in TMDs~\cite{Cao2012,Xiao2012}, the spin of the optically excited electron-hole pair (and hence the photon polarization) is linked to the valley index. Therefore, we define
\beq \label{eq:Hg}
\hat{H}_{\rm{ph-mat}} 
= g \underset{\k\q\sg}{\sum} \ed_{\sg, \frac{\q}{2}+\k} \hd_{\sg, \frac{\q}{2}-\k} \ch_{\sg,\q} +
h.c. ,
\eeq
with $g$ being the bare light-matter coupling constant. Since this also corresponds to a contact interaction, it requires renormalization~\cite{Mead1991}. For simplicity, we take the ultraviolet cutoff (on the relative electron-hole momentum) to be the same as for the electron-hole interactions, since it is governed by the same length scale, i.e., the lattice spacing. Note that, in writing Eq.~\eqref{eq:Hg}, we have applied the rotating wave approximation which is justified since we are working with cavity photon energies which are comparable with the band gap, which in turn greatly exceeds all other relevant energy scales in the problem.

\subsection{Foundations of the diagrammatic approach}

To describe the intervalley polariton-electron scattering and trion resonance, we use a microscopic description of the exciton-polariton~\cite{Li2021,Levinsen2019Microscopic}.  Here, we briefly review results which are important for the diagrammatic formulation of scattering involving polaritons. For further details, we refer the reader to  
Appendices \ref{ehTmatrix} and \ref{Polaritons}.

We start by considering the bare electron/hole Green's function. This is also called the propagator, and takes the form
\beq
G^{\rm{e/h}}_{\sg}(\p,E)=\Frac{1}{E-\ep^{\rm{e/h}}_{\p}+i0}.
\eeq
This describes the free motion of an electron/hole with momentum $\p$ and spin $\sg$ in the absence of interactions and light-matter coupling. The energy pole corresponds to the dispersion. The imaginary infinitesimal $+i0$ shifts the poles of the Green's function slightly into the lower half plane such that $G$ corresponds to a retarded Green's function, as appropriate for a few-body scattering problem. In the following, we will always be assuming that the energy carries a positive imaginary part, that is, all Green's functions will be understood to be retarded.

\begin{figure*}
\includegraphics[width=0.76\linewidth]{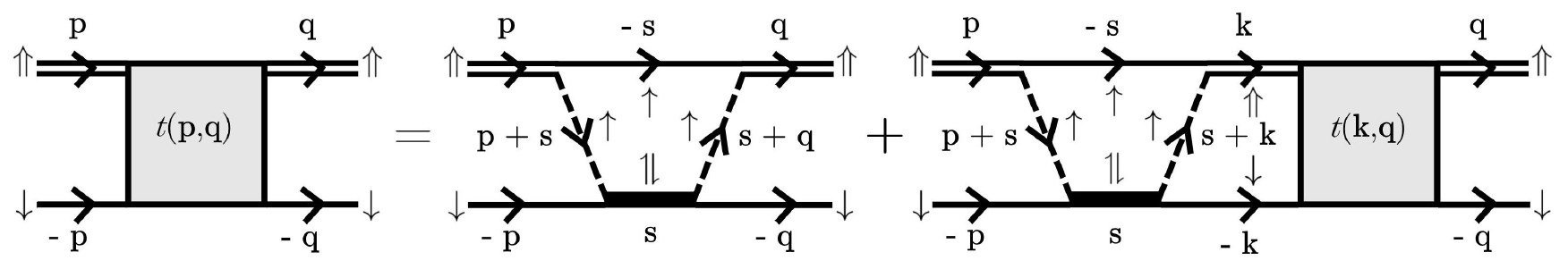}
\caption[system]{\label{STMFig}
Diagrammatic representation of the three-body equation \eqref{STM}, from which we obtain the polariton-electron scattering \textit{T} matrix (shaded square) as a function of incoming and outgoing momenta $\pm\p$ and $\pm\q$. The polariton formed by the optically active electron-hole pair and cavity photon is drawn as a double line, the electron propagator is drawn as a single line, the hole propagator as a dashed line, and the exciton propagator as a solid bar. Note that the incoming and outgoing polariton and electron propagators factor out in the three-body equation.}
\end{figure*}

The sum of all possible processes featuring the repeated scattering of an electron and a hole leads to the electron-hole \textit{T} matrix. In the absence of coupling to light, the \textit{T} matrix at total momentum $\p$ and energy $E$ takes the form
\beq \label{ExcitonPorpagator}
T_0(\p,E)=\Frac{-2\pi/m_{\rm{r}}}{\ln({\Frac{\ep^{\rm{X}}_{\p} -E}{\eb}})},
\eeq
where $\ep^{\rm{X}}_{\p} = \p^2/2m_{\rm{X}}$ is the exciton kinetic energy with exciton mass $m_{\rm{X}}=m_{\rm{e}}+m_{\rm{h}}$. The renormalization of the electron-hole contact interaction is carried out in detail in Appendix~\ref{ehTmatrix}.

To obtain the polariton properties within our model~\eqref{eq:Ham}, we first consider the propagator of the cavity photon in the presence of the semiconductor. The fully renormalized photon propagator is given by~\cite{Li2021}
\beq \label{PhotonPropagator}
D_{\sg}(\p,E)=\frac{1}{
E-\w-\ep^{c}_{\p} +\frac{\Omega^2}{\eb} \big{[} \ln (\frac{\ep^{\rm{X}}_{\p} -E}{\eb}) \big{]} ^{-1}
}.
\eeq
For details, see Appendix \ref{Polaritons}. Here, the cavity photon frequency is shifted from its bare value $\omega$ due to the interaction with the semiconductor medium. As a result, it can be related to the physical photon-exciton detuning $\delta$ via $\w=\delta-\eb+\frac{\Omega^2}{2\eb}$~\cite{Li2021}. The polariton dispersion is found by solving for the poles of the propagator, and thus satisfies the transcendental equation
\begin{align}
    E=\w+\ep^{\rm{c}}_{\p} -\frac{\Omega^2}{\eb} \big{[} \ln (\frac{\ep^{\rm{X}}_{\p} -E}{\eb}) \big{]} ^{-1}.
\end{align}
As long as $\Omega,\delta\lesssim\varepsilon_\mathrm{X}$, there are two solutions of this equation, corresponding to the lower (-) and upper (+) polariton dispersions $E_\pm(\p)$. Furthermore, the matter-part of the polariton is responsible for mediating the interaction in polariton-electron scattering. The exciton fraction, or squared exciton Hopfield coefficient, is given by 
\beq\label{ExcitonFraction}
X^{2}_{\pm} (\p)= \frac1{1+
\frac{\eb}{\Omega^2} 
\ln(\frac{\ep^{\rm{X}}_{\p}-E_{\pm}
(\p)}{\eb})^2
 (\ep^{\rm{X}}_{\p}-E_{\pm}(\p))}
.
\eeq

Finally, we note that, in the presence of coupling to light, the electron-hole \textit{T} matrix is modified to~\cite{Li2021}
\beq \label{PolaritonPropagator}
P_{\sg}(\p,E)=\frac{-2\pi/m_{\rm{r}}}{
\ln(\frac{\ep^{\rm{X}}_{\p}-E}{\eb}) +
\frac{\Omega^2}{\eb} 
(E-\w-\ep^{\rm{c}}_{\p})^{-1}
}.
\eeq
This can be thought of as a polariton propagator, and indeed it has the same pole structure as the photon propagator in Eq.~\eqref{PhotonPropagator}. 
As discussed in Appendix \ref{Polaritons}, it includes all the potential scattering processes between the electron and hole, along with the repeated transformation of the electron-hole pair into a photon and vice versa.

\section{Polariton-electron scattering} \label{PolaritonElectronScattering}

We now consider the intervalley polariton-electron scattering and associated trion resonance, as illustrated in Fig.~\ref{TrionFig}. We use a diagrammatic technique that allows us to straightforwardly include all contributions to the interaction. Our formulation is similar to the celebrated Skorniakov and Ter-Martirosian equation, first introduced in the context of neutron-deuteron scattering in nuclear physics~\cite{Skorniakov1957}, and since adopted to the description of the scattering of an atom and a diatomic molecule in the context of ultracold atomic gases~\cite{Petrov2003STM,Brodsky2006,Levinsen2006,helfrich2011three,Ngampruetikorn2013}.  Formally, our calculation is also closely related to the case of intravalley polariton-electron scattering~\cite{Li2021PRL,Li2021}, where the electron occupies the same valley as the electron inside the polariton. The primary difference is that in our case the two electrons are distinguishable which enables them to bind to a hole to form a trion, while the Pauli principle suppresses trion formation for indistinguishable electrons  (see, e.g., Ref.~\cite{Tiene2022} for a detailed discussion of this point). We remind the reader that while we refer specifically to polariton-electron scattering in the following, our results also apply trivially to polariton-hole scattering, of relevance in hole-doped semiconductor microcavities.

The central object of the polariton-electron scattering is the $T$ matrix, which is the (appropriately normalized) sum of all diagrams contributing to the scattering process. Since all terms are of the same order of magnitude, it is not possible to simply calculate each term in the sum individually and then sum them all. Instead, Fig.~\ref{STMFig} shows how the (unnormalized) sum can be related to itself via the integral equation:
\begin{align}
t(\p,\q,E) = & B(\p,\q,E) \nonumber \\
& +\underset{\k}{\sum} B(\p,\k,E) P_{\uparrow}(\k,E-\ep^{\rm{e}}_{\k}) t(\k,\q,E) ,
\label{STM}\end{align}
similar to the Lippmann-Schwinger equation of two-body scattering. Here, the first term on the right hand side is the sum of all diagrams which have the polariton and electron as external legs but never inside the diagram---we refer to this term as the irreducible exchange process $B$. As indicated in the figure, we will be working in the center of mass frame where the incoming (outgoing) polariton and electron have momenta $\pm\p$ ($\pm\q$). The associated energies of the incoming and outgoing electrons are taken to be $\epsilon_\p^{\rm{e}}$ and $\epsilon_\q^{\rm{e}}$ while that of the polaritons is $E-\epsilon_\p^{\rm{e}}$ and $E-\epsilon_\q^{\rm{e}}$ with the total energy $E$. In the continuum limit, Eq.~\eqref{STM} constitutes a Fredholm integral equation in the first argument of $t$, which we solve numerically by projecting onto the different partial waves and discretizing the integrals using the Gauss-Legendre quadrature~\cite{numericalrecipes}. More details are in Appendix~\ref{PartialWave}.

\begin{figure*}[th]
\includegraphics[width=0.99\linewidth]{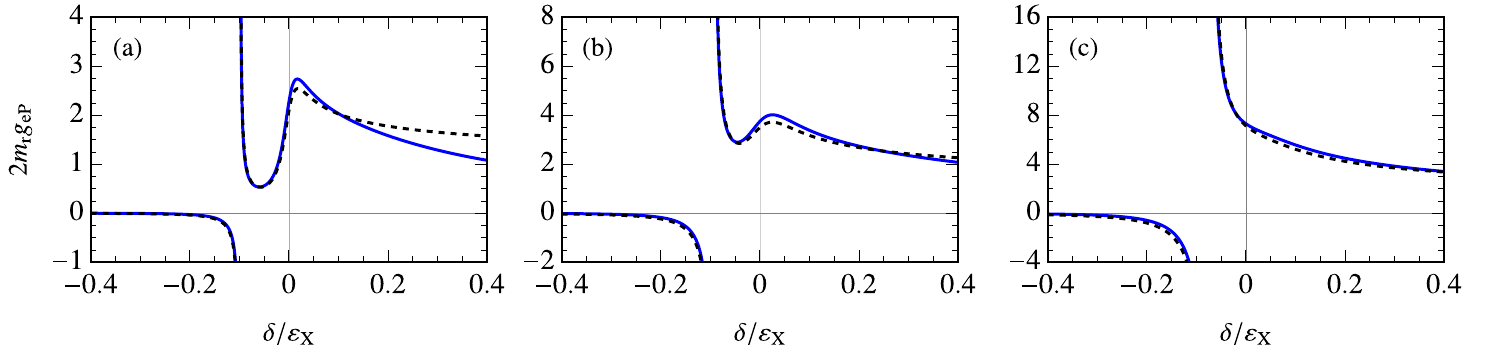}
\caption[system]{\label{Tdetuning}
The polariton-electron coupling constant $g_{\rm{eP}} = T_s(0)$ as a function of detuning for equal masses ($m_\mathrm{e}=m_\mathrm{h}$), and $\Omega/\eb =$ (a) 0.01, (b) 0.025 (corresponding approximately to monolayer MoSe$_2$, MoS$_2$ and WSe$_2$) and (c) 0.05 (corresponding approximately to monolayer WS$_2$). We show both the results of our full numerical calculation (solid blue) and the analytic approximation in Eq.~\eqref{AnalyticT} (dashed black). Note that the scale of $g_{\rm{eP}}$ increases from (a) to (c).
}
\end{figure*}

The irreducible exchange process $B$ is represented by the first diagram on the right-hand side of the three-body equation shown in Fig.~\ref{STMFig}. This shows how the hole is first transferred from the incoming spin-up polariton to the optically inactive electron, where it interacts and propagates internally as a \replyadd{dark} exciton. Following this, the hole is then transferred back to the spin-up polariton, completing the hole-exchange process. Specifically, we have
\beq
\begin{split} \label{eq:irredB}
B(\p,\q,E) = & \sum_{\s} 
\chi(\p+\frac{\s}{2})
G^{\rm{h}}_{\uparrow} ( \p+\s, E-\ep^{\rm{e}}_{\p}-\ep^{\rm{e}}_{\s} ) \\
& \chi(\s+\frac{\p}{2})
T_0 ( \s,E-\ep^{\rm{e}}_{\s} ) 
\chi(\s+\frac{\q}{2}) \\
& G^{\rm{h}}_{\uparrow} ( \s+\q, E-\ep^{\rm{e}}_{\s}-\ep^{\rm{e}}_{\q} ) \chi(\q+\frac{\s}{2})  .\\
\end{split}
\eeq 
This irreducible exchange process is then iterated an arbitrary number of times to obtain the $T$ matrix, interspersed with propagation of the polariton and electron.

In Eq.~\eqref{eq:irredB}, $\chi$ is a form factor that we have introduced to tune the ratio between the exciton and the trion binding energies. In the absence of coupling to light and for $\chi=1$,  the three-body equation~\eqref{STM} features a trion bound state with energy $E_\mathrm{T}=-2.39\varepsilon_\mathrm{X}$~\cite{Brodsky2005,Pricoupenko2010} (obtained as a pole of $t$ at $E<-\varepsilon_\mathrm{X}$ and zero momentum). This corresponds to a binding energy of $|E_\mathrm{T} + \eb| = 1.39$ which is much larger than that expected in 2D semiconductors, where $E_\mathrm{T}\simeq-1.1\varepsilon_\mathrm{X}$~\cite{Thilagam97,Sergeev2001,Courtade2017}. The discrepancy is primarily due to how we have neglected the electron-electron repulsion. 
To effectively introduce this repulsion, we note that the exchange of a hole between the polariton and exciton physically requires the two electrons to come into close proximity. We can thus mimic the repulsion through the use of form factors in the three-body exchange diagrams, which amounts to suppressing their large-momentum (short-range) contribution, while leaving the original two-body theory in Sec.~\ref{TheoreticalFramework} unchanged. 
Such an approach has previously proven successful in nuclear theory~\cite{Haidenbauer1984Separable,Strobel1968Separable,Grinyuk2009Separable}, cold atoms~\cite{Krzysztof2004,Jona2008,Laird2018}, and light-matter coupled systems~\cite{Levinsen2019}. 

Specifically, we take the form factor $\chi$ to act on the relative motion between the two interacting charges at the four interaction vertices of $B$. For the results presented in the main text, we utilise a Gaussian form factor
\beq \label{GaussianFormAngle}
\chi (\p) = e^{-p^2/\Lambda_{3}^{2}} ,
\eeq
with the parameter $\Lambda_3=0.8234 a_{\rm{X}}$ tuned to fix $E_\mathrm{T}\simeq-1.1\varepsilon_\mathrm{X}$. In Appendix~\ref{FormfactorAppendix} we presents results for several different functional forms of the form factor, with the results being almost completely independent of the precise choice. This highlights the universality of our results, with the features of the polariton-electron scattering dominated by the interplay between strong coupling to light and the existence of a trion, rather than by the precise form of the electronic interactions. 

We now turn from the trion bound states to quantifying the strength of scattering between electrons and polaritons. At momenta characteristic of polaritons, i.e., up to around the polariton inflection point, the scattering is dominated by the rotationally symmetric $s$-wave channel, and we therefore consider only this. Conservation of energy requires that the incoming and outgoing relative momenta of the particles are equal. Furthermore, the energy $E$ of a physical scattering process must be on-shell, implying that it equals the total energy of the polariton and electron, $E=E_-(\p)+\epsilon_\p^{\rm{e}}$. Thus, similarly to the case of intravalley polariton-electron scattering~\cite{Li2021}, the normalized on-shell \textit{T} matrix takes the form
\beq \label{NumericT}
T_s (p)  =  Z_{-}(\p)|X_{-}(\p)|^{2} t_s (\p,\p,E_{-} (\p)+ \ep^{\rm{e}}_{\p}),
\eeq
where the overall normalization $Z_{\pm} (\p) |X_{-}(\p)|^{2}= (2\pi/m_{\rm{r}}) |E_{\pm} (\p) - \ep^{\rm{X}}_{\p}| |X_{-}(\p)|^{2}$ is the residue of the polariton propagator at its energy pole. From the $T$ matrix, we obtain the long-wavelength polariton-electron coupling constant
\begin{align}
    g_{\rm{eP}} = T_s(0). 
\end{align}
This is the coupling constant that one would use as a starting point for a mean-field description of a many-body system of polaritons and electrons.

Figure \ref{Tdetuning} shows the polariton-electron coupling constant as a function of photon-exciton detuning. The results are calculated for three different strengths of the Rabi coupling, with the results for $\Omega=0.025\eb$ ($\Omega=0.05\eb$) corresponding approximately to the monolayers MoSe$_2$, MoS$_2$ and WSe$_2$ (WS$_2$). In Fig.~\ref{Tdetuning}, the trion pole can be seen for each light-matter coupling: as the detuning is changed from negative to positive we cross a critical detuning where the collision energy equals the trion energy, i.e., $E_-(0)=E_\mathrm{T}$, and where consequently the on-shell \textit{T} matrix diverges.  
For detunings where the polariton energy is below the trion, the resulting $g_{\rm{eP}}$ is negative, indicating a strong attraction between the polariton and electron.  Conversely, as we increase the detuning past the critical point, we see positive values of $g_{\rm{eP}}$, indicating a strong repulsion. At positive detuning, another resonance-like structure is seen for the weaker Rabi couplings. This is also present in same valley polariton-electron scattering~\cite{Li2021}, and originates from an interplay between the strong energy dependence of the underlying exciton-electron scattering and the strong detuning dependence of the excitonic Hopfield coefficient. Generically, we see that the trion resonance broadens as the light-matter coupling increases, and that the resulting coupling constant is very large around zero detuning where most experiments are performed.

\begin{figure*}[ht]
\includegraphics[width=0.99\linewidth]{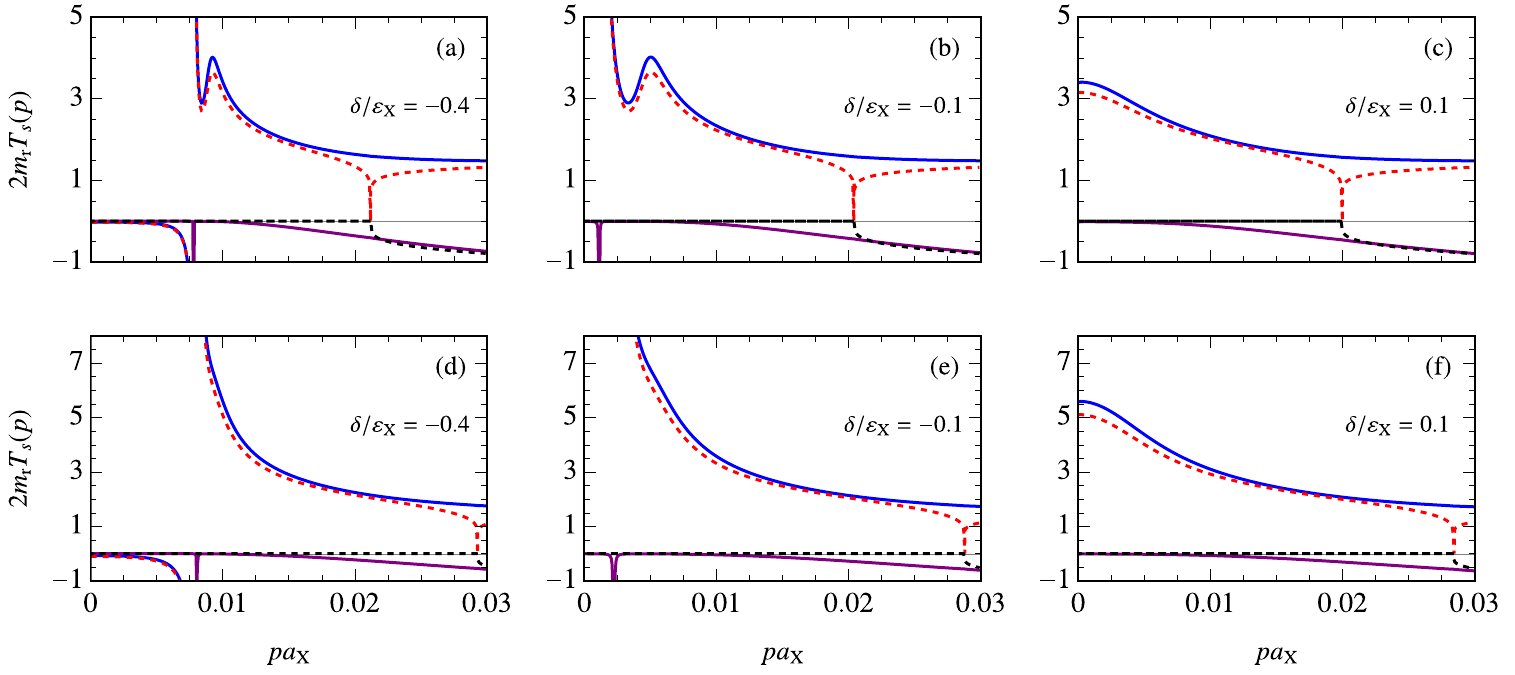}
\caption[system]{\label{Tmomentum}
The real (solid blue, dashed red) and imaginary parts (solid purple, dashed black) of our calculated polariton-electron \textit{T} matrix $T_s (p)$ (solid lines) and the analytic approximation in Eq.~\eqref{AnalyticT} (dashed lines). Here we use TMD parameters with $\Omega/\eb=0.025$ (top row) and 0.05 (bottom row).
}
\end{figure*}

To better understand the behavior of polariton-electron scattering we can consider an analytic approximation of the \textit{T} matrix. This is based on the idea that while the strong coupling to light shifts the collision energy, the actual interaction process is dominated by the underlying interactions between the charges. Hence, polariton-electron scattering can be understood as off-shell exciton-electron scattering, an idea that has also been successfully employed to describe polariton-polariton interactions~\cite{Bleu2020,Li2021PRBPP} and intravalley polarion-electron interactions~\cite{Li2021,Li2021PRL}. At low collision energy compared with the exciton binding energy, the exciton-electron scattering takes the universal form of low-energy short-range interactions~\cite{AdhikariAJP86}, leading to
\beq \label{AnalyticT}
  T_{s} (p) \simeq X_{-}(p)^{2} \Frac{2\pi}{m_{\rm{eX}}}\frac1
  {\ln [
  -\ep_{\rm{eX}}/(E_{-}(p)+\ep^{\rm{e}}_{p}+\eb)]},
\eeq
which reduces to Eq.~\eqref{AnalyticTintro} in the introduction in the limit $p\to0$. Here,  the constant $\epsilon_{\rm{eX}}=0.1\eb$ is approximately the trion binding energy (with possibly minor corrections at very strong light-matter coupling), such that the pole of the \textit{T} matrix approximately corresponds to the experimentally observed trion binding energy. In Fig.~\ref{Tdetuning} we plot the analytical \textit{T} matrix as dashed black lines, and we see that this approximate \textit{T} matrix agrees extremely well. The agreement with the universal low-energy form of polariton-electron scattering further emphasizes the universal nature of our results.

In fact, the non-zero value of $g_\mathrm{eP}$ is a remarkable consequence of the broken Galilean invariance in the light-matter coupled system. Indeed, 2D scattering theory predicts that the scattering amplitude should vanish at zero momentum~\cite{landau2013quantum} for any short-range interaction (such as the exciton-electron potential). However, in the light-matter coupled system this only happens for $p\lesssim \exp(-m_\mathrm{r}/m_\mathrm{c})a_\mathrm{X}^{-1}$, a momentum scale that is so small that it is only relevant in systems much larger than the size of the universe~\cite{Bleu2020}. Hence, in practice our results for $g_\mathrm{eP}$ apply when $p\ll a_\mathrm{X}^{-1}$.

In Fig.~\ref{Tmomentum} we plot the real and imaginary parts of the \textit{T} matrix as a function of relative momentum for different values of detuning and Rabi coupling, still focusing on parameters relevant to monolayer TMDs (top row: MoSe$_2$, MoS$_2$ and WSe$_2$, bottom row WS$_2$). In Fig.~\ref{Tmomentum} (a) and (b) we see that the trion resonance occurs at finite collision momentum. The origin is the strongly momentum-dependent polariton dispersion below the inflection point, with the resonance corresponding to a very high degree of accuracy to when the polariton dispersion crosses the trion energy, i.e., to the condition $E_-(p)=E_\mathrm{T}$. In panel (c), the detuning is such that we are above the trion resonance for all collision momenta (refer to Fig.~\ref{Tdetuning}(b)). For this value of the Rabi coupling, we see that we can still observe a second resonance-like feature above the trion resonance. 
On the other hand, for larger light-matter coupling (lower panels) the trion pole broadens and dominates the finite momentum behavior, and the resonance-like peak is absorbed into the trion pole. In both cases, we find that the trion resonance also shows up as a narrow peak in the imaginary part of the full \textit{T} matrix (purple line). However, this peak is finite and therefore much smaller than the real part, as discussed in Appendix \ref{PartialWave}.     

We also compare our finite-momentum results with the analytic approximation in Eq.~\eqref{AnalyticT}. This is again seen to capture the behavior of the trion pole and resonance-like structure extremely well. The only point where it fails is close to where the collision energy matches the exciton energy, i.e., when $E=-\varepsilon_\mathrm{X}$, where we find a fictitious onset of the imaginary part. This is due to the analytic formula neglecting the continuum of states that exist below the exciton in the light-matter coupled system, and it is associated with the argument of the logarithm in Eq.~\eqref{AnalyticT} becoming negative. This feature can in principle be cured by introducing a small photon linewidth (i.e., taking $E_-(p)\to E_-(p)+i\Gamma$), as discussed in Appendix \ref{app:impart}. It is also absent in a full two-body $T$ matrix calculation that treats the exciton as tightly bound but includes the finite photon mass~\cite{Bleu2020}.

\begin{figure}
\includegraphics[width=0.99\linewidth]{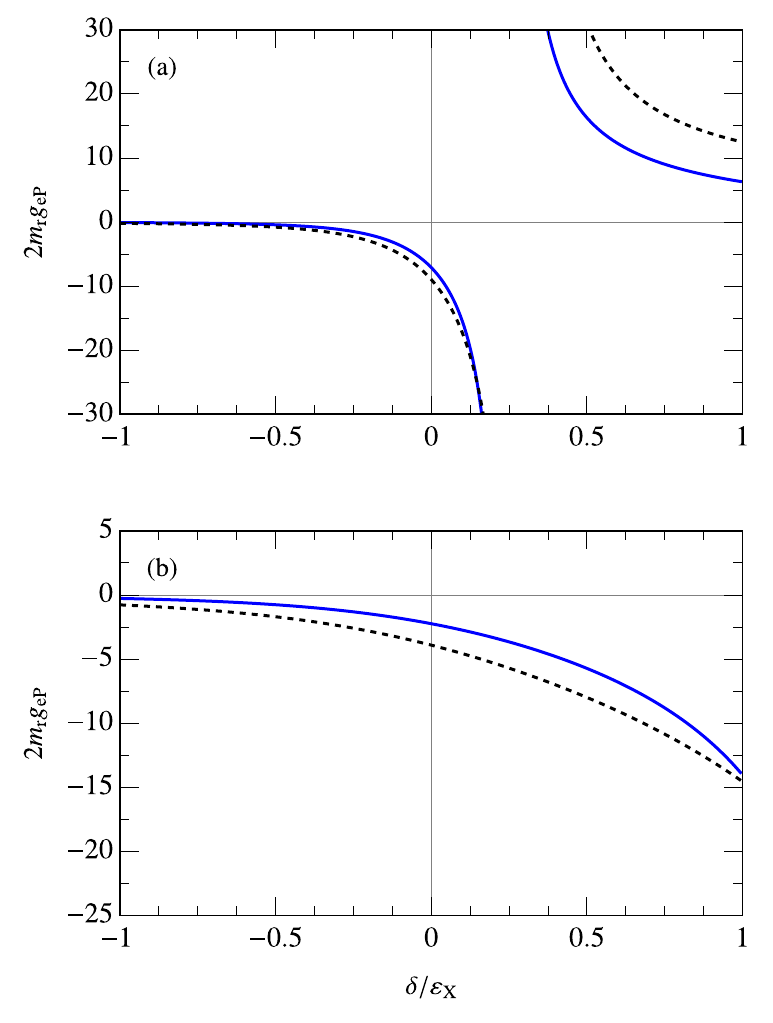}
\caption[system]{\label{Tdetuninggaas}
The polariton-electron coupling constant $g_{\rm{eP}}$ as function of detuning for $\Omega/\eb =$ 0.2 and 0.5 are shown in panel (a) and (b), respectively. Here, we use GaAs parameters, with $m_{\rm{e}}=0.067m_0$ and $m_{\rm{h}}=0.45m_0$, where $m_0$ is the vacuum electron mass. We show both the result of our numerical calculation (solid blue) and the analytic approximation in Eq.~\eqref{AnalyticT} (dashed black), where we take $\epsilon_{\rm{eX}}=0.1\eb$. 
}
\end{figure}

Aside from monolayer TMDs, our results also apply to conventional 2D quantum well semiconductor microcavities such as GaAs. These naturally feature larger light-matter coupling relative to their exciton binding energy, even in the case of a microcavity containing only a single quantum well. We show our results for $g_\mathrm{eP}$ for parameters corresponding to a single GaAs quantum well in Fig.~\ref{Tdetuninggaas}(a) and for a system with larger Rabi coupling in panel (b). Here, we take the trion energy to be $E_{\rm T}=-1.1\eb$ as in the case of monolayer TMDs. For a single quantum well, we predict a broad resonance at positive detuning, similar to those in monolayer TMDs. This should in principle be observable by methods similar to those employed in Ref.~\cite{SidlerNatPhys16}. Once we go to larger Rabi coupling, the trion resonance shifts to prohibitively large positive detunings. This explains why the trion resonance is not observed in doped GaAs quantum well microcavities, since typically these feature multiple quantum wells to enhance the light-matter coupling, which suppresses the trion resonance around zero detuning.

\begin{figure*}
\includegraphics[width=0.99\linewidth]{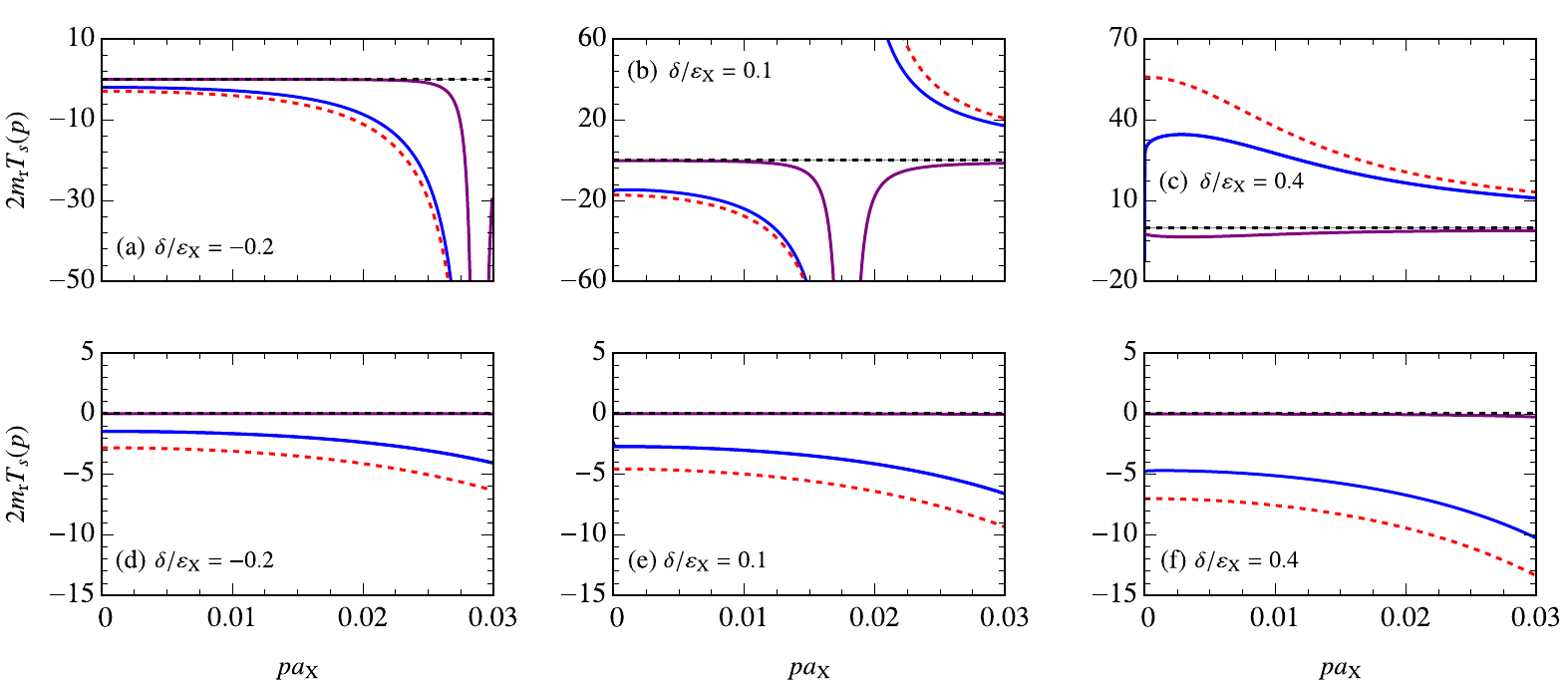}
\caption[system]{\label{Tmomentumgaas}
The real (solid blue, dashed red) and imaginary parts (solid purple, dashed black) of the full \textit{T} matrix $T_s (p)$ and the analytic approximation in Eq.~\eqref{AnalyticT} using $\epsilon_{\rm{eX}}=0.1\eb$. Here we use GaAs parameters with $\Omega/\eb=0.2$ (top row) and 0.5 (bottom row).
}
\end{figure*}

In Fig.~\ref{Tmomentumgaas} we show results for GaAs for polariton-electron scattering at finite momentum. Here, we again find that the resonance is most likely to be observed in a microcavity containing a single quantum well. We also find that the imaginary part of the scattering amplitude can be significant close to the trion resonance, although still substantially smaller than the real part.

Finally, we briefly discuss the role played by multiple semiconductor layers in a microcavity. It is well known that the effective Rabi coupling scales as $\sqrt{N}$, where $N$ is the number of layers. Less appreciated is the fact that the effective pairwise interactions involving polaritons (i.e., polariton-polariton, polariton-exciton or polariton-electron scattering) is reduced by a factor $1/N$ due to the polariton being spread over multiple layers~\cite{Bleu2020}. For instance, our results in the lower row of Figs.~\ref{Tdetuninggaas} and \ref{Tmomentumgaas} correspond roughly to a Rabi coupling relevant to six layers, and therefore the effective coupling constant and $T$ matrix should be divided by this factor if one were to use these parameters as inputs into a mean-field calculation of a polariton-electron mixture. In spite of this, it appears realistic to observe the trion resonance in microcavities containing a few doped TMD monolayers.

\section{Concluding remarks} \label{ConcludingRemarks}

To conclude, we have used a diagrammatic theory to investigate intervalley polariton-electron scattering. Our theory allowed us to sum all diagrams contributing to the scattering, under the approximation of strongly screened electronic interactions. In particular, we observed the coupling of the scattering process to the trion, resulting in a drastic enhancement of the polariton-electron interaction strength when the polariton energy is tuned to that of the trion. A simple analytic approximation provided further evidence of the universality of our results, i.e., their independence of the precise form of the electronic interactions. Overall, our results indicate a large degree of tunability in polariton-electron interactions, since the polariton energy can be tuned by changing the cavity frequency, or even potentially by applying a magnetic field~\cite{GaAsLargeRabi,Pietka2017,Laird2022}.

It would be interesting to extend our theory to the case of longer-range electronic interactions, such as the Coulomb or Rytova-Keldysh potentials characterizing atomically thin 2D semiconductors. Technically, this could be done by using the analytic Coulomb \textit{T} matrix originally derived by Schwinger~\cite{Schwinger2005,CombescotPRX2017}. Of particular importance would be the investigation of whether interactions between Rydberg polaritons and electrons could also be resonantly enhanced, a precise description of which is beyond the current approach based on screened electronic interactions. 

Our theory provides a microscopic foundation for the polariton-electron interactions in effective theories of polaron polaritons in charge-doped semiconductor microcavities~\cite{Baeten2015,SidlerNatPhys16}. In particular, our analytic approximations in Eqs.~\eqref{AnalyticTintro} and \eqref{AnalyticT} can be used as a starting point for $T$-matrix based theories of polarons and polariton-electron mixtures in a range of semiconductor heterostructures, including moir\'e superlattices in twisted bilayers~\cite{Shimazaki2020,Campbell2022} which are only beginning to be explored. 

\acknowledgments 
We acknowledge useful discussions with Olivier Bleu and Francesca Maria Marchetti. This research was supported by the Australian Research Council
Centre of Excellence in Future Low-Energy Electronics Technologies
(CE170100039). JL and MMP are also supported through the Australian Research
Council Future Fellowships FT160100244 and FT200100619, respectively.

\appendix \label{Appendix}

\section{Electron-hole \textit{T} matrix}  
\label{ehTmatrix}

Here, we derive the electron-hole \textit{T} matrix~\cite{LevinsenBook15}. We first write the Lippmann-Schwinger equation
for the $T$ matrix at total momentum $\q$ and energy $E$ as  
\beq \label{BareT0}
\begin{split}
T_0 (\q,E)  = & -V_0 + V_0\sum_\k^\Lambda \frac{1}{E-\ep^{\rm{X}}_{\q}-\ep^{\rm{r}}_{\k}+i0} V_0 -\cdots \\
 = & -V_0 + V_0 \Pi(E-\ep^{\rm{X}}_{\q}) V_0 -  \cdots \\
 = & \frac{-1}{V_0^{-1} + \Pi(E-\ep^{\rm{X}}_{\q}) } ,
\end{split}
\eeq
where we have defined
\begin{align}
    \Pi(E)=\sum_\k^\Lambda \frac1{E-\epsilon_\k^\mathrm{r}+i0}.   
\end{align}
The exciton (i.e., the center of mass) kinetic energy is $\ep^{\rm{X}}_{\q} = \q^2 / 2m_{\rm{X}}$ with $m_{\rm{X}}=m_{\rm{e}}+m_{\rm{h}}$, 
and the kinetic energy of the relative motion is $\ep^{\rm{r}}_{\k} = \k^2 / 2m_{\rm{r}}$. In 2D, the integral in $\Pi(E) \sim \ln (\Lambda)$ is divergent as the cutoff $\Lambda \rightarrow \infty$.  Thus, in order to have physically meaningful results in the limit $\Lambda\to\infty$ we must have $V_0^{-1} \sim -\Pi(E-\ep^{\rm{X}}_{\q})$, corresponding to carrying out a renormalization procedure. Using the condition that the \textit{T} matrix at $\q=0$ should have a pole at the exciton energy $E=-\eb$, we obtain
\beq
V_{0}^{-1} = - \Pi(-\eb) .
\eeq
Substituting back into Eq.~\eqref{BareT0} gives the renormalized \textit{T} matrix
\beq
 T_0 (\q,E) = \frac{-1}{ \Pi(E-\ep^{\rm{X}}_{\q}) - \Pi(-\eb 
 ) } =  \frac{-2\pi/m_{\rm{r}}}{
\ln (\frac{\ep^{\rm{X}}_{\q}-E}{\eb}-i0) 
 } ,
\eeq
where we have taken the limit $\Lambda\to\infty$.

An important quantity is the residue of the \textit{T} matrix at $E=-\eb$ for zero momentum, i.e.,
\beq
Z_X = \frac{2\pi}{m_{\rm r}} \eb .
\eeq
This is the square of the normalization of the exciton wavefunction.

\section{Exciton-polaritons at finite momentum}\label{Polaritons}

We now go through the theoretical description of exciton polaritons within the model~\eqref{eq:Ham}. The single-polariton problem in a model containing photons and their coupling to electrons and holes interacting via a Coulomb potential was first solved in Ref.~\cite{Levinsen2019Microscopic}. In the present case of contact electron-hole interactions, the polariton problem was first addressed in Ref.~\cite{Hu2020}. Here, we use the renormalization scheme of Ref.~\cite{Li2021PRL}, which has the advantage of being fully analytic. We now briefly review this approach.

Considering only states consisting of the photon or an electron-hole pair, the dressed photon propagator $D(\q,E)=\bra{0} \ch_{\q} (E-\hat{H})^{-1} \chd_{\q} \ket{0}$ can be expanded in powers of the light-matter interaction to give the Dyson equation (for simplicity, we suppress the spin index throughout this appendix)
\beq
\begin{split}
D(\q,E) & = D_0 (\q,E) + D_0 (\q,E) \Sigma(\q,E) D(\q,E) \\
& = \frac{1}{D_0 (\q,E)^{-1} -\Sigma(\q,E) } .
\end{split}
\eeq
Here, the free photon propagator is $D_0(\q,E)=(E-\w-\ep^{\rm{c}}_{\q}+i0)^{-1}$ and the photon self energy $\Sigma$ consists of two parts 
\beq
 \Sigma(\q,E) = g^2 \Pi (E-\ep^{\rm{X}}_{\q}) + g^2 \Pi (E-\ep^{\rm{X}}_{\q}) T(\q,E) \Pi ( E-\ep^{\rm{X}}_{\q}),
 \label{eq:photonSE}
\eeq
where the first term corresponds to the photon creating an electron-hole pair that recombines to form a photon, while the second term corresponds to those processes where the electron-hole pair interact following their creation.
As discussed in Appendix \ref{ehTmatrix}, the integral $\Pi(E-\ep^{\rm{X}}_{\q})$ with the momentum cutoff $\Lambda$ is divergent, and in order to get a finite result we must have $g \sim 1/\ln(\Lambda)$. This implies that the momentum cutoff for light-matter coupling and electron-hole interaction behave in the same manner, i.e., $g \sim V_0$ as $\Lambda\to\infty$. The simplest renormalization scheme is to take the cutoff on the light-matter coupling to be the same as for the electron-hole interaction, as in Ref.~\cite{Li2021}. Within this scheme, we renormalize the bare light-matter coupling constant such that $g^2 \Pi^2 (E
) = \Omega^2 / Z_{\rm{X}} $ as $\Lambda \rightarrow \infty$, with the relation between the coupling constants being $g=\Omega \frac{V_0}{\sqrt{Z_{\rm{X}}}}$. 
This results in the first term of Eq.~\eqref{eq:photonSE} going to zero. The resulting renormalized dressed photon propagator is 
\beq \label{DressedPhotonPropagatorA}
D(\q,E) = \frac{1}{
E-\w-\ep^{\rm{c}}_{\q}+\frac{\Omega^2}{\eb}
\big{[}
\ln(\frac{\ep^{\rm{X}}_{\q}-E-i0}{\eb})
\big{]}^{-1}+i0
} .
\eeq
The bare cavity photon energy $\w$ is also renormalized, with its relation to the physical detuning $\delta$ between the cavity photon and the 1$s$ exciton being 
\beq
\w=\delta-\eb+\frac{\Omega^2}{2\eb}.
\eeq
This is due to the optically active semiconductor medium shifting the bare cavity photon energy.

The photon and exciton fractions, or the squared Hopfield coefficients, are found using the unitary pole expansion. The photon fraction $C^{2}_{\pm} (p)$ is obtained by calculating the residue of the dressed photon propagator Eq.~\eqref{PhotonPropagator}. For a Green's function $G(E)=(E-\Sigma(E))^{-1}$ with a pole at energy $E_0$, the residue by the pole expansion is given by $Z=(1-\frac{d \mathrm{Re}\Sigma(E)}{dE} \big{|}_{E=E_0} )^{-1}$. By expanding the photon propagator in Eq.~\eqref{DressedPhotonPropagatorA} around its poles, the photon fraction is found to be
\beq\label{PhotonFraction}
C^{2}_{\pm} (\p)= \bigg{(}
1+\frac{\Omega^2}{\eb} 
\frac{1}{\left[\ln(\frac{\ep^{\rm{X}}_{\p}-E_{\pm}(\p)}{\eb})\right]^2
 (\ep^{\rm{X}}_{\p}-E_{\pm}(\p))}
\bigg{)}^{-1} .
\eeq
Using the photon fraction, we can extract the exciton fraction from the normalization condition: $C^{2}_{\pm}(\p)+X^{2}_{\pm}(\p)=1$.
 
The electron-hole \textit{T} matrix in the presence of light-matter coupling, which can be thought of as a polariton propagator $P$, is derived based on the matter component of the polariton capable of interacting with a third particle. In the absence of light-matter coupling, $P$ should reduce to the electron-hole \textit{T} matrix $T_0$. Following Ref.~\cite{Li2021}, and in condensed form ignoring the energy and momentum variables,
\beq
\begin{split}
P & = T_0+ g^2 (
T_0 \Pi D + T_0 \Pi D \Pi T_0 + D + D \Pi T_0
) \\
& = T_0 +T_0 D g^2 \Pi^2 \\
& = \frac{1}{T_{0}^{-1} - D_0 g^2 \Pi^2 } ,
\end{split}
\eeq
where in going from the first to the second line we took the limit $\Lambda\to\infty$, cancelling terms that vanish in this limit.
Thus, within our model, only the diagrams which begin and end with the formation of the electron and hole are capable of interacting with the third particle in polariton-electron scattering.
This leads to the polariton propagator
\beq
P_{\sg}(\p,E)=\frac{-2\pi/m_{\rm{r}}}{
\ln(\frac{\ep^{\rm{X}}_{\p}-E-i0}{\eb}) +
\frac{\Omega^2}{\eb} 
(E-\w-\ep^{\rm{c}}_{\p})^{-1}
+i0}.
\eeq

\section{Partial-wave decomposition and numerical solution of the scattering equations} \label{PartialWave}
In a similar manner to the partial-wave decomposition of a two-body \textit{T} matrix and potential, we project the three-body equation onto its partial waves. We first write the partial-wave decomposition of the whole polariton-electron scattering \textit{T} matrix and irreducible exchange process:
\beq \label{PartailWaveDecomp}
\begin{split}
& t (\p,\q,E) = \overset{\infty}{\underset{l=0}{\sum}} (2-\delta_{l0}) \cos(l \theta_{\p\q}) t_l (p,q,E) , \\
& B (\p,\q,E) = \overset{\infty}{\underset{l=0}{\sum}} (2-\delta_{l0}) \cos(l \theta_{\p\q}) B_l (p,q,E) .
\end{split}
\eeq
Here, $\delta$ is the Kronecker delta and $\theta_{\p\q}=\theta_{\p}-\theta_{\q}$ is the difference between the angles associated with the vectors $\p$ and $\q$. By inverting Eq.~\eqref{PartailWaveDecomp}, we have
\beq
\begin{split}
& t_l (p,q,E) = \int_{0}^{2\pi} \frac{d\theta_{\p\q}}{2\pi} \cos(l \theta_{\p\q}) t(\p,\q,E) , \\
& B_l (p,q,E) = \int_{0}^{2\pi} \frac{d\theta_{\p\q}}{2\pi} \cos(l \theta_{\p\q}) B(\p,\q,E) .
\end{split}
\eeq

Now we write the three-body equation Eq.~\eqref{STM} in integral form by taking the continuum limit, i.e., converting the sum over momentum to an integral using 
\beq
\underset{\k}{\sum} = \int^{\infty}_{0} \frac{kdk}{2\pi}
\int^{2\pi}_{0} \frac{d\theta_{\k\q}}{2\pi},
\eeq
where the angle of momentum $\k$ is taken with reference to $\q$. Then, we insert the expressions in Eq.~\eqref{PartailWaveDecomp} for $t$ and $B$ into Eq.~\eqref{STM}. Focusing on the angular integration in the second term on the right hand side of Eq.~\eqref{STM}, and integrating with respect to $\theta_{\k\q}$ gives
\beq \label{angularintegral}
\begin{split}
\overset{\infty}{\underset{n,m=0}{\sum}} &
(2-\delta_{n0}) (2-\delta_{m0}) \times \\
& \int^{2\pi}_{0} 
\cos{(n\theta_{\p\k})} \cos{(m\theta_{\k\q})}
\frac{d\theta_{\k\q}}{2\pi} \\
& = \overset{\infty}{\underset{m=0}{\sum}} 
(2-\delta_{m0}) \cos{(m\theta_{\p\q})} .
\end{split}
\eeq
Here, $\theta_{\p\k}$ and $\theta_{\k\q}$ equal $\theta_\p-\theta_\k$ and $\theta_\k-\theta_\q$, respectively. Now we apply the integral operator $\int^{2\pi}_{0} \cos(l\theta_{\p\q}) [.] \frac{d\theta_{\p\q}}{2\pi}$ to the left and right hand sides of Eq.~\eqref{STM} to isolate the $l^{th}$ partial-wave equation, giving
\beq \label{STMpartialwave}
\begin{split}
t_{l}(p,q,E) = & B_{l}(p,q,E) + \\
& \int^{\infty}_{0} B_{l}(p,k,E) P_{\sg}(k,E-\ep^{\rm{e}}_{k}) t_{l}(k,q,E) \frac{kdk}{2\pi} .
\end{split}
\eeq

Solving for the partial-wave exchange process term $B_l$ is done in a similar manner to the kernel of the scattering equation. For convenience we define the partial-wave hole propagator
\beq \label{holeprop}
\begin{split}
& g^{\rm{h}}_{\sg l} (p,q,E) = \int_{0}^{2\pi} \frac{d\theta_{\p\q}}{2\pi}
 \cos(l\theta_{\p\q})
g^{\rm{h}}_{\sg} (\p,\q,E) \\
& = \int_{0}^{2\pi} \frac{d\theta_{\p\q}}{2\pi} 
\cos(l\theta_{\p\q}) \times \\
& \bigg{(} E-\frac{p^2}{2m_{\rm{e}}}-\frac{q^2}{2m_{\rm{e}}}- 
\frac{p^2+q^2+2pq\cos(\theta_{\p\q})}{2m_{\rm{h}}}
\bigg{)}^{-1},
\end{split}
\eeq
and its inverse
\beq \label{Ghpartialwave}
g^{\rm{h}}_{\sg} (\p,\q,E) = \overset{\infty}{\underset{l=0}{\sum}} (2-\delta_{l0}) \cos(l \theta_{\p\q}) g^{\rm{h}}_{\sg l} (p,q,E) .
\eeq
Ignoring for now the form factor $\chi$ (see discussion in Appendix~\ref{FormfactorAppendix} about how this is included in practice), the partial-wave exchange process in integral form is 
\beq \label{ExchangeDecompose}
\begin{split}
B (\p,\q,E) = & 
\int_{0}^{\infty} \frac{s ds}{2\pi}
\int_{0}^{2\pi} \frac{d\theta_{\s\q}}{2\pi} 
\times \\
& g^{\rm{h}}_{\sg} (\p,\s,E)
T_0 (\s,E-\ep^{\rm{e}}_{\s}) 
g^{\rm{h}}_{\sg} (\s,\q,E) .
\end{split}
\eeq
Now we use Eq.~\eqref{PartailWaveDecomp} and Eq.~\eqref{Ghpartialwave} and insert for $B$ and $g^{\rm{h}}_{\sg}$ in Eq.~\eqref{ExchangeDecompose}. This produces a very similar set of equations to the kernel of the three-body equation above. Using the same integration procedure as in Eq.~\eqref{angularintegral}, and then applying the same integral operator $\int^{2\pi}_{0} \cos(l\theta_{\p\q}) [.] \frac{d\theta_{\p\q}}{2\pi}$ to $B$, we isolate the partial-wave contribution
\beq  \label{ExchangeTermPartialwave}
\begin{split}
B_{l} (p,q,E) = & 
\int_{0}^{\infty} \frac{s ds}{2\pi}
 g^{\rm{h}}_{\sg l} (p,s,E)
T_0 (s,E-\ep^{\rm{e}}_{s}) 
g^{\rm{h}}_{\sg l} (s,q,E) .
\end{split}
\eeq

\begin{figure*}[th]
\includegraphics[width=0.86\linewidth]{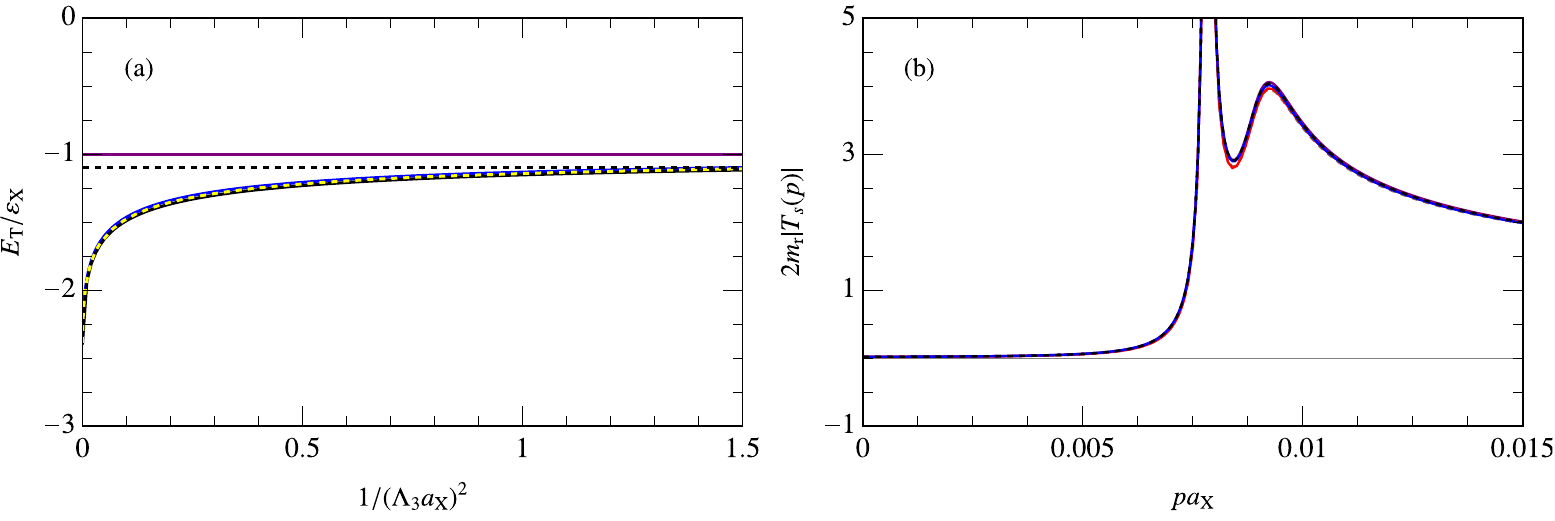}
\caption[system]{\label{AppendixCombinedFig}
(a) Trion energy $E_{\rm{T}}$ as a function of the form factor parameter $1/(\Lambda_3 a_{\rm{X}} )^2$ for the Gaussian (solid blue), Yamaguchi (solid black) and Yamaguchi squared form factors (dashed yellow). The exciton energy $-\eb$ (solid purple) is shown for reference. We also show the realistic trion energy $E_{\rm{T}}=-1.1\eb$ (dashed black): the intersection between the calculated trion energy with this line fixes the form factor parameter $\Lambda_3$. 
(b) Absolute value of the full \textit{T} matrix for $\Omega/\eb=0.025$ and $\delta/\eb=-0.4$. Calculations using the Gaussian, Yamaguchi and Yamaguchi squared form factors are shown using solid blue, purple, and red. Their respective angular integration that do not apply the approximation Eq.~\eqref{eq:forms} are shown using dashed black, brown, and gray. The lines are essentially indistinguishable, providing evidence for the universality of our results.
}
\end{figure*}

We solve for the partial-wave \textit{T} matrix using the principal value method, as the polariton and exciton propagators have poles that are integrated over in the three-body equation~\eqref{STMpartialwave} and in the hole-exchange process~\eqref{ExchangeTermPartialwave}. In the following, we demonstrate how we treat the pole in the kernel of the 
scattering equation \cite{BedaquePRC1998}, with the pole in the hole-exchange process carried out in a similar manner. The issue is that the polariton propagator has a simple pole when the energy is on-shell, i.e., when the collision energy matches the sum of the single-particle energies: $E=E_{-}(q)+\epsilon_q^{\rm{e}}$. Thus $P(k,E_{-}(q)+\ep^{\rm{e}}_{q}-\ep^{\rm{e}}_{k}) \rightarrow \infty$ as $k \rightarrow q$. We treat this using the Sokhotski–Plemelj theorem
\beq
\frac{1}{k-q-i 0} = \frac{\mathcal{P}}{k-q} +i\pi \delta(k-q), 
\eeq
where $\mathcal{P}$ indicates the principal part.
This allows us to write the three-body equation Eq.~\eqref{STMpartialwave} with $E=E_{-}(q)+\epsilon_q^{\rm{e}}$ as 
\beq
\begin{split}
t_l (p,q) = [1-i\pi\alpha(q) & 
t_l (q,q)]  B_l (p,q) + \\
& \mathcal{P} \int^{\infty}_{0} B_l (p,k) P(k) t_l (k,q) \frac{kdk}{2\pi} ,
\end{split}
\eeq
where we have suppressed the energy dependence of the various terms.
Here, we define
\beq
\begin{split}
\alpha(q) & =-\underset{k \rightarrow q}
{\lim}
\frac{k(k-q)}{2\pi} P(k,E_{-}(q)+\ep^{\rm{e}}_{q}-\ep^{\rm{e}}_{k}) \\
& = \frac{q}{2\pi} \frac{
|X_{-}(q)|^{2} Z_{-} (q)
}{
\partial (E_{-}(q)+\ep^{\rm{e}}_{q})/ \partial q 
}
\end{split}
\eeq
and
\beq
Z_{\pm} (q)=\frac{2\pi}{m_{\rm{r}}} |E_{\pm} (q) - \epsilon_{q}^{\rm{X}}| .
\eeq
For convenience we define the \textit{T} matrix such that 
\beq \label{DecomposedT}
t_l (p,p) = \frac{1}{
K_{l}^{-1} (p,p) + i \pi \alpha(p)
} .
\eeq
Here, we can see that the real and imaginary parts are separated, and the quantity $K_l$ is found from the equation
\beq
K_{l}(p,q) = B_{l}(p,q) + 
 \mathcal{P} \int^{\infty}_{0} B_{l}(p,k) P_{\sg}(k) K_{l}(k,q) \frac{kdk}{2\pi} .
\eeq

As seen in Fig.~\ref{Tmomentum}, at negative detuning the imaginary part of the $T$ matrix has a sharp and narrow peak as the collision energy matches the trion energy. To study this peak, we isolate the imaginary part of the full \textit{T} matrix using Eq.~\eqref{DecomposedT}. As the real part of the \textit{T} matrix tends to infinity 
\beq \label{ImaginaryTmatrixPeak}
\text{Im} ( t_l (p,p) ) = - \frac{1}{\pi \alpha(p)} .
\eeq
The value of $\alpha$ is proportional to the exciton fraction, which is very small for negative detunings, thus resulting in the sharp (but finite) feature.

\section{Form factors and exchange process}\label{FormfactorAppendix} 

\begin{table}[h]
\renewcommand{\arraystretch}{1.9}
\begin{tabular}{|c|c|c|}
\hline
\multicolumn{2}{|c|}{
Functional form of $\chi (p)$
}
&\multicolumn{1}{|c|}{
Critical $\Lambda_3$ value
}\\
\hline

\multicolumn{1}{|p{3cm}|}{ 
\centering Gaussian
} 
& \multicolumn{1}{|p{2.5cm}|}{
\centering 
$
   \text{exp} (- p^2 / \Lambda_{3}^{2} )
$
}
& \multicolumn{1}{|p{2cm}|}{
\centering 
$
   0.8234 a_{\rm{X}}^{-1}
$
}\\
\hline

\multicolumn{1}{|p{3cm}|}{ 
\centering \text{Yamaguchi}
} 
& \multicolumn{1}{|p{2.5cm}|}{
\centering 
$
    (1+p^2/\Lambda_{3}^{2})^{-1} 
$
}
& \multicolumn{1}{|p{2cm}|}{
\centering 
$
   0.7316 a_{\rm{X}}^{-1}
$
}\\
\hline

\multicolumn{1}{|p{3cm}|}{ 
\centering Yamaguchi Squared
} 
& \multicolumn{1}{|p{2.5cm}|}{
\centering 
$
    (1+ p^2/ 2 \Lambda_{3}^{2})^{-2}
$
}
& \multicolumn{1}{|p{2cm}|}{
\centering 
$
   0.7768 a_{\rm{X}}^{-1}
$
}\\
\hline

\end{tabular}
\caption{\label{ParameterTable}
Table of the form factor functions used and their respective critical values of $\Lambda_3$ for which $E_\mathrm{T}=-1.1\varepsilon_\mathrm{X}$.  
}
\end{table}

Here, we discuss the different form factors $\chi(\p)$ we have used in our calculations. These reduce the strength of processes where the hole is exchanged between the two electrons, thus mimicking electron-electron repulsion. Their functional form and the critical value of their respective parameters $\Lambda_3$ used to reproduce the observed trion energy of $E_{\rm{T}}=-1.1\eb$ are tabulated in Table~\ref{ParameterTable}. 

For the results shown in Table~\ref{ParameterTable} and in the main text, we have made the approximation of taking the \textit{s}-wave projection of the form factor, i.e.,
\beq
(\p+\q)^2 \to p^2+q^2.\label{eq:forms}
\eeq
While this simplification is not exact, it offers a substantial numerical advantage since it is necessary for the decoupling into separate partial waves discussed in Appendix~\ref{PartialWave}. Furthermore, it is an extremely good approximation since the coupling to higher angular momentum channels is strongly suppressed for momenta $\lesssim 1/a_\mathrm{X}$.

To illustrate the universality of our results, Fig.~\ref{AppendixCombinedFig}(a) demonstrates that the three form factors behave almost identically while solving for the trion energies $E_{\rm{T}}$ as a function of $\Lambda_3$. Furthermore, Fig.~\ref{AppendixCombinedFig}(b) shows our results for the momentum-dependent scattering using all three types of form factors (with the \textit{s}-wave approximation in Eq.~\eqref{eq:forms}) as well as for the form factors without this approximation (where we solve for the full angle-dependent $T$ matrix, which is numerically much more expensive). We see that the results are essentially identical, which provides strong evidence that our polariton-electron scattering results are truly universal.

\section{Imaginary part in the analytic approximation to polariton-electron scattering}
\label{app:impart}

\begin{figure}
\includegraphics[width=0.99\linewidth]{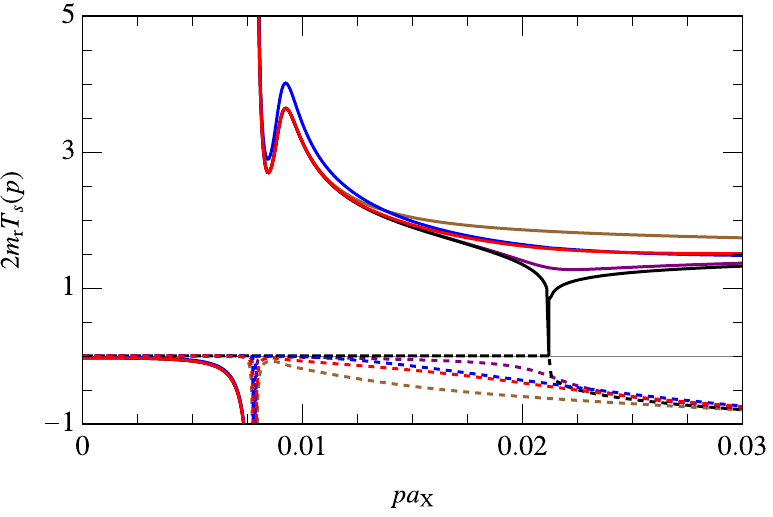}
\caption[system]{\label{TiGamma}
The real (solid) and imaginary parts (dashed) of the polariton-electron \textit{T} matrix $T_s (p)$ for $\Omega/\eb=0.025$ and $\delta=-0.4$, with equal electron and hole masses. The numerical calculation (blue), the analytic approximation Eq.~\eqref{AnalyticT} with $\Gamma=0$ (black), $\Gamma=1\times10^{-4}\eb$ (purple), $\Gamma=4\times10^{-4}\eb$ (red) and $\Gamma=1\times10^{-3}\eb$ (brown) are shown. 
}
\end{figure}

In the analytic \textit{T} matrix, Eq.~\eqref{AnalyticT}, the spurious sharp onset of the imaginary part disappears as soon as we introduce a small photon linewidth in the energy, i.e. $E_{-} (p) \rightarrow E_{-} (p) + i \Gamma$, which always exists in real experiments. This has been illustrated for different values of $\Gamma$ in Fig.~\ref{TiGamma}. We see that all values of $\Gamma$ used reproduce the qualitative features of the numerical results very well, with $\Gamma=4\times10^{-4}\eb$ providing the best match.

\bibliography{e_P_scattering_refs.bib}

\end{document}